\documentclass[conference]{IEEEtran}
\IEEEoverridecommandlockouts

\usepackage{cite}
\usepackage{amsmath,amssymb,amsfonts}
\usepackage{graphicx}
\usepackage{textcomp}
\usepackage{xcolor}
\usepackage{url}
\usepackage[hidelinks]{hyperref}
\usepackage{xurl}
\begin{document}

\title{The Environmental Impact of AI Servers and Sustainable Solutions}

\author{
\IEEEauthorblockN{Rusheen Patel\IEEEauthorrefmark{1}, Nikhil Mahalingam\IEEEauthorrefmark{1}, Aadi Patel\IEEEauthorrefmark{1}}
\IEEEauthorblockA{\IEEEauthorrefmark{1}Department of Electrical \& Computer Engineering, Rutgers University--New Brunswick\\
Course: Sustainable Energy (14:332:402)\\
December 16, 2025}
}

\maketitle

\begin{abstract}
The rapid expansion of artificial intelligence (AI) has significantly increased the electricity, water, and carbon demands of modern data centers, raising sustainability concerns. This study evaluates the environmental footprint of AI server operations and examines feasible technological and infrastructural strategies to mitigate these impacts. Using a literature-based methodology supported by quantitative projections and case-study analysis, we assessed trends in global electricity consumption, cooling-related water use, and carbon emissions. Projections indicate that global data center electricity demand may increase from approximately 415 TWh in 2024 to nearly 945 TWh by 2030, with AI workloads accounting for a disproportionate share of this growth \cite{ref1}. In the United States alone, AI servers are expected to drive annual increases in water consumption of 200--300 billion gallons and add 24--44 million metric tons of CO$_2$-equivalent emissions by 2030 \cite{ref4,ref5}. The results show that the design of the cooling system and the geographic location influence the environmental impact as strongly as the efficiency of the hardware. Advanced cooling technologies can reduce cooling energy by up to 50\%, while location in low-carbon and water-secure regions can cut combined footprints by nearly half. In general, the study concludes that sustainable AI expansion requires coordinated improvements in cooling efficiency, renewable energy integration, and strategic deployment decisions.
\end{abstract}


\section{Introduction}
Artificial intelligence has become a cornerstone of modern computation, enabling advances in fields ranging from healthcare and finance to transportation and scientific research. However, rapid scaling of AI models has led to a substantial increase in the size, density, and energy intensity of data centers. High-performance AI servers require continuous power and cooling, placing a growing strain on electrical grids, water resources, and climate targets worldwide. Unlike traditional enterprise computing, AI workloads operate at sustained high utilization levels, significantly amplifying both energy consumption and thermal output. As a result, the environmental footprint of AI-focused data centers differs fundamentally from that of conventional facilities. In regions with constrained infrastructure, this rapid expansion has begun to expose vulnerabilities in grid capacity, water availability, and emission reduction pathways. These pressures are expected to intensify as AI adoption accelerates across both public and private sectors. Addressing these challenges requires a system-level understanding of how AI infrastructure interacts with energy and resource networks.

The goal of this project is to quantify the environmental impacts associated with AI server operations and to evaluate practical, scalable solutions that can reduce these impacts. Specifically, this study focuses on the growth of electricity demand, carbon emissions, cooling-related water consumption, and the influence of geographic location. Previous work from industry and research institutions, including global energy demand assessments by the International Energy Agency (IEA) \cite{ref1}, national data center energy analyzes by Lawrence Berkeley National Laboratory (LBNL) \cite{ref2}, and infrastructure case studies such as Microsoft’s Project Natick \cite{ref8}, provide the basis for this analysis.

This paper is organized as follows. Section~II describes the methodology and data sources used. Section~III presents results related to the impacts of energy, water, and carbon on the growth of AI servers. Section~IV reviews sustainable solutions, including advanced cooling, integration of renewable energy, and alternative siting strategies. Section~V concludes with key findings, limitations, and directions for future work.

\section{Methodology}
This study employs a literature-based research methodology to assess the environmental impacts of AI servers and evaluate sustainability solutions. Due to the nature of operational data from hyperscale data centers, publicly available reports, peer-reviewed studies, and industry case analyzes were used as primary data sources.

Quantitative projections of electricity demand were drawn from global and U.S.-specific assessments by the IEA \cite{ref1} and Goldman Sachs \cite{ref3}. The energy usage trends of the historical and baseline data center were informed by reports from Lawrence Berkeley National Laboratory \cite{ref2}. The impacts of water consumption and carbon emissions were analyzed using national estimates and regional case studies, including academic modeling of AI infrastructure footprints \cite{ref5,ref6}. Qualitative evaluations of emerging solutions, such as liquid cooling and underwater data centers, were based on documented industry deployments, including Meta’s StatePoint liquid cooling system \cite{ref9} and Microsoft’s Project Natick \cite{ref8}.

Key assumptions include steady AI workload growth through 2030 and average grid carbon intensities based on current regional mixes. These assumptions are consistent with previous large-scale modeling studies and allow for a comparative evaluation of mitigation strategies.

\subsection{Limitations and Uncertainty Considerations}
Although this study relies on well-established projections and peer-reviewed models, several sources of uncertainty should be acknowledged. First, future AI workload growth rates depend on advances in model efficiency, algorithmic optimization, and hardware specialization, all of which may alter long-term energy demand trajectories. Second, regional electricity grid carbon intensities are expected to evolve over time as renewable penetration increases, which may reduce emissions relative to present-day estimates. Third, water consumption values vary significantly depending on the design of the local cooling system, climate conditions, and regulatory constraints, creating uncertainty when extrapolating national-scale impacts from regional data.

Additionally, much of the operational data from hyperscale data centers remain proprietary, requiring reliance on public disclosures and modeling assumptions. Despite these limitations, the use of multiple independent data sources and consistent trends across studies suggests that overall conclusions regarding the magnitude and drivers of AI-related environmental impacts remain robust.

\section{Results}

\begin{figure}[!t]
\centering
\includegraphics[width=\columnwidth]{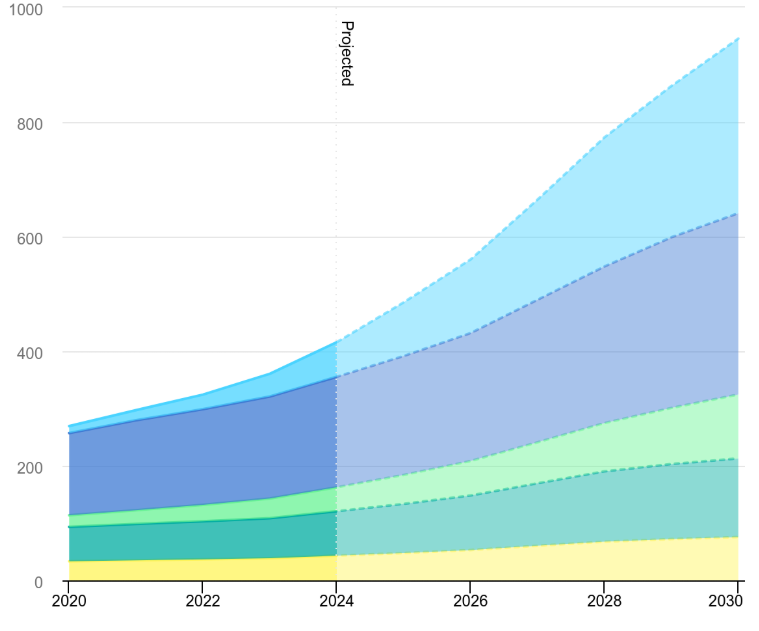}
\caption{Projected increase in global data center electricity demand (post-2024 surge driven by AI/compute workloads), with cooling contributing an increasing share of total energy use \cite{ref1}.}
\label{fig:dc_power}
\end{figure}

\begin{figure}[!t]
\centering
\includegraphics[width=\columnwidth]{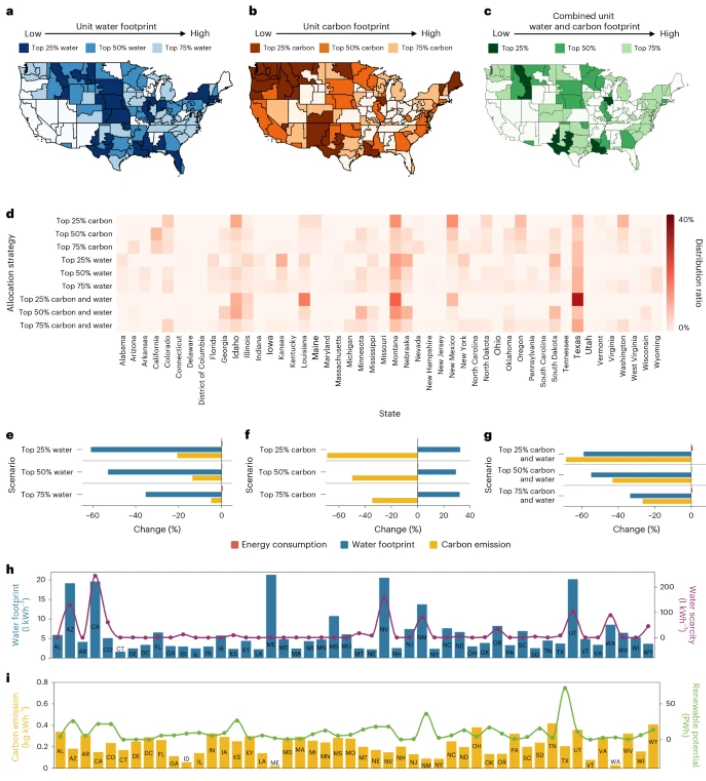}
\caption{Geographic variation in unit water footprint, unit carbon footprint, and combined environmental impact of AI data center operations across U.S.\ states, highlighting the dependence of outcomes on siting decisions \cite{ref5,ref6}.}
\label{fig:geo_impacts}
\end{figure}

\subsection{Growth in Global Electricity Demand}
The growth of AI workloads is driving a rapid increase in global data center electricity consumption. In 2024, data centers consumed approximately 415 TWh of electricity, representing approximately 1.5\% of total global demand. By 2030, this figure is projected to increase to nearly 945 TWh, or roughly 3\% of global electricity consumption, corresponding to an annual growth rate of approximately 15\% \cite{ref1}. This rate significantly exceeds the growth of global electricity demand and indicates that AI-driven computation will represent a major future load on power systems.

From a systems perspective, the projected increase in AI-driven electricity demand is comparable in scale to the current annual electricity consumption of mid-sized industrialized nations. This comparison highlights the potential grid-level implications of AI expansion, including increased peak demand, transmission congestion, and the need for accelerated grid modernization. Without targeted efficiency improvements and demand-side management, AI data centers could significantly influence future grid planning and infrastructure investment decisions.

This rapid post-2024 surge in electricity demand, led primarily by AI and compute workloads, is illustrated in Fig.~\ref{fig:dc_power}, which also highlights the growing contribution of cooling systems to total data center energy use \cite{ref1}.

\subsection{Water Consumption from AI Servers}
In addition to electricity demand, AI servers require water resources for cooling. Between 2024 and 2030, AI servers in the United States alone are projected to consume between 200 and 300 billion gallons of water annually. These estimates isolate AI workloads and do not include other data center operations, underscoring the unique and growing burden imposed by AI infrastructure. The demand for water is particularly concerning in arid and semi-arid regions, where the expansion of data centers is increasingly competing with municipal and agricultural needs \cite{ref4}.

\subsection{Carbon Emissions Associated with AI Expansion}
Carbon emissions are also expected to increase sharply as the AI infrastructure scales. Depending on the intensity of the carbon electricity that supplies data centers, AI servers in the United States could generate an additional 24--44 million metric tons of CO$_2$-equivalent emissions annually by 2030. This wide range reflects differences in grid composition and emphasizes the importance of low-carbon electricity sources for AI deployments \cite{ref5}. Without targeted mitigation strategies, AI growth could significantly hinder national and global de-carbonization efforts.

\subsection{Impact of Location and Grid Mix}
Geographic location and grid composition were found to influence environmental outcomes as strongly as hardware efficiency. Having AI data centers in favorable regions versus unfavorable regions can reduce carbon emissions by up to 49\% or increase them by up to 90\%. Similarly, water footprints can decrease by approximately 52\% or increase by more than 350\% depending on location. These regional disparities in unit water footprint, unit carbon footprint, and combined environmental impact across the United States are illustrated in Fig.~\ref{fig:geo_impacts} \cite{ref5,ref6}.

\subsection{Implications for Grid Planning and Resource Allocation}
Beyond absolute increases in electricity demand, the rapid growth of AI workloads has important implications for grid planning and resource allocation. AI data centers operate with high load factors and limited flexibility, which can exacerbate peak demand conditions and reduce the effectiveness of traditional demand-side management strategies. As a result, regions experiencing rapid AI infrastructure growth may require accelerated investments in transmission capacity, substation upgrades, and local generation resources to maintain grid reliability.

In addition, the coupling of electricity demand with water-intensive cooling systems creates interdependencies between energy and water infrastructure. In water-stressed regions, increased cooling demand can restrict future data center expansion or shift environmental burdens to alternative water sources, such as reclaimed or desalinated water. These system-level interactions suggest that AI infrastructure planning must increasingly consider the effects of the energy–water nexus, rather than optimizing electricity consumption or water use in isolation.

\section{Sustainable Solutions}

\subsection{Advanced Cooling Technologies}
Cooling systems represent a major source of both energy and water consumption in AI data centers. Transitioning from traditional air and evaporative cooling to advanced techniques such as liquid cooling, direct-to-chip cooling, and immersion cooling can significantly improve efficiency. Depending on climate and system design, these technologies can reduce cooling-related electricity consumption by approximately 40--50\% and water usage by 20--90\% \cite{ref7}. Large-scale deployments, such as Meta’s StatePoint liquid cooling system, demonstrate the technical feasibility and scalability of these approaches \cite{ref9}.

\subsection{Renewable Energy and Microgrids}
De-carbonization of the electrical supply is essential for reducing AI-related carbon emissions. An effective strategy is the deployment of on-site or local microgrids that incorporate solar, wind, and battery storage. In favorable conditions, these systems can reduce operational emissions by 50--80\% by providing a significant fraction of AI energy demand with low-carbon electricity \cite{ref3}. In addition, renewable microgrids improve grid resilience and reduce dependence on carbon-intensive peak generation.

Beyond emissions reduction, renewable-powered microgrids can also improve the operational reliability of AI data centers. As AI workloads increasingly support mission-critical services, interruptions caused by grid instability or extreme weather events pose growing risks. On-site generation combined with battery storage enables data centers to maintain continuous operation during grid outages while reducing reliance on diesel backup generators. In regions with high renewable variability, intelligent load scheduling and energy storage coordination can further align AI computation with periods of low-carbon electricity availability, improving both sustainability and system resilience.

\subsection{Geographic Siting and Alternative Infrastructure Models}
Strategic geographic location is one of the most impactful solutions to reduce the environmental footprint. Relocating AI data centers from water-scarce regions such as California, Nevada, and Arizona to water-secure regions with cleaner grids, such as parts of the Midwest and Great Plains, can substantially reduce both water and carbon impacts \cite{ref6}. Alternative infrastructure models are also being explored. Microsoft’s Project Natick demonstrated that underwater data centers can take advantage of ocean temperatures for passive cooling, reducing cooling energy requirements and improving reliability \cite{ref8}. Although not universally applicable, such approaches highlight innovative pathways toward sustainable deployment.

\subsection{Economic and Infrastructure Trade-Offs}
Although advanced cooling systems, renewable energy integration, and strategic positioning provide clear environmental benefits, they also introduce economic and infrastructure trade-offs that influence real-world adoption. Advanced cooling technologies such as liquid immersion and direct-to-chip cooling require a higher upfront capital investment, specialized facility retrofitting, and increased operational expertise. These costs can pose barriers to retrofitting existing data centers, particularly those built around traditional air-cooled designs.

Similarly, renewable-powered microgrids and on-site energy storage systems require significant initial investment in generation capacity, batteries, and power electronics. Although long-term operational savings and emission reductions can offset these costs, economic feasibility often depends on local electricity prices, regulatory incentives, and utility interconnection policies. In regions with limited renewable resources or restrictive grid regulations, data center operators may face challenges in achieving high renewable penetration without relying on off-site power purchase agreements.

Infrastructure constraints also play a role in geographical location decisions. Although water-secure and low-carbon regions offer environmental advantages, they may lack existing high-capacity transmission infrastructure, fiber connectivity, or proximity to the main user bases. As a result, optimal environmental positioning must balance sustainability objectives with latency requirements, reliability considerations, and economic constraints. These trade-offs highlight the need for coordinated planning between data center operators, utilities, and policy makers to enable environmentally sustainable AI growth on-scale.

\subsection{Policy and System-Level Considerations}
Although technological solutions play a central role in reducing the environmental impacts of AI servers, policy and regulatory frameworks are equally important. Incentives for the adoption of renewable energy, water-efficient cooling technologies, and low-carbon data center placement can accelerate the deployment of sustainable AI infrastructure. Grid interconnection policies, carbon pricing mechanisms, and reporting requirements for energy and water in data centers can further improve transparency and accountability. In addition, standardized sustainability benchmarks for AI infrastructure would enable consistent comparison across operators and regions. Public disclosure requirements can also encourage competition toward lower environmental footprints. Finally, aligning regulatory incentives with long-term climate targets can reduce the risk of short-term optimization at the expense of sustainability. At the international level, harmonized standards could prevent the relocation of environmentally intensive AI infrastructure to regions with weaker regulations. Public–private partnerships may further lower barriers to adoption by sharing risk and accelerating deployment of sustainable technologies. Regulatory clarity can also reduce uncertainty for long-term infrastructure investments. Together, these measures help ensure that sustainability considerations are embedded throughout the AI infrastructure lifecycle rather than treated as secondary constraints.

At the system level, coordination between data center operators, utilities, and policy makers is necessary to ensure that AI growth aligns with grid capacity planning and long-term de-carbonization goals. As AI workloads increasingly support critical services, incorporating sustainability metrics into infrastructure planning will be essential to balance performance, reliability, and environmental responsibility.

From a long-term perspective, integrating sustainability metrics into AI infrastructure decision-making can become as important as performance and cost considerations. As AI increasingly underpins critical services, policy frameworks that encourage environmentally responsible deployment can help balance innovation with resilience and resource stewardship.

\section{Conclusions}
This project examined the environmental impacts of AI servers and found that rapid AI expansion is driving significant increases in electricity demand, water consumption, and carbon emissions. The results indicate that cooling technology, energy sourcing, and geographic location are critical determinants of sustainability outcomes. Although no single solution is sufficient on its own, a combination of advanced cooling, the integration of renewable energy, and strategic placement can dramatically reduce AI’s environmental footprint. Together, these findings highlight that infrastructure-level decisions can be as impactful as improvements in computational efficiency. Moreover, early investment in sustainable design can prevent long-term environmental lock-in as AI systems scale. Finally, coordinated planning across energy, water, and digital infrastructure is necessary to achieve meaningful reductions at scale.

The limitations of this study include the reliance on publicly available data and projections rather than operational measurements. Future work should focus on lifecycle carbon assessments, economic analyzes of retrofitting existing facilities, and region-specific evaluations of cooling performance. Continued research and policy support will be essential to ensure that AI innovation progresses in parallel with environmental sustainability.

These findings underscore the importance of addressing sustainability concerns early in the design and deployment of AI infrastructure. As AI adoption accelerates across sectors, incorporating environmental considerations into engineering, policy, and planning decisions will be critical to ensuring long-term scalability and societal benefit.

\begin{sloppypar}

\end{sloppypar}


\begin{thebibliography}{9}

\bibitem{ref1}
International Energy Agency, ``Energy and AI: Energy demand from artificial intelligence,'' IEA, 2024. [Online]. Available: \url{https://www.iea.org/reports/energy-and-ai/energy-demand-from-ai}

\bibitem{ref2}
Lawrence Berkeley National Laboratory, ``2024 LBNL data center energy usage report,'' LBNL, 2024. [Online]. Available: \url{https://eta.lbl.gov/publications/2024-lbnl-data-center-energy-usage-report}

\bibitem{ref3}
Goldman Sachs, ``AI to drive 165\% increase in data center power demand by 2030,'' Goldman Sachs Insights, 2024. [Online]. Available: \url{https://www.goldmansachs.com/insights/articles/ai-to-drive-165-increase-in-data-center-power-demand-by-2030}

\bibitem{ref4}
``Thirsty for power and water, AI-crunching data centers sprout across the West,'' \emph{And the West}, Apr. 8, 2025. [Online]. Available: \url{https://andthewest.stanford.edu/2025/thirsty-for-power-and-water-ai-crunching-data-centers-sprout-across-the-west/}

\bibitem{ref5}
Y.~Zhang \emph{et al.}, ``Environmental impacts of large-scale AI data centers,'' \emph{arXiv preprint} arXiv:2509.21312, 2025. [Online]. Available: \url{https://arxiv.org/abs/2509.21312}

\bibitem{ref6}
J.~Patterson \emph{et al.}, ``Carbon and water impacts of artificial intelligence infrastructure,'' \emph{Nature Sustainability}, 2025. [Online]. Available: \url{https://www.nature.com/articles/s41893-025-01681-y}

\bibitem{ref7}
Nortek Data Center Cooling, ``Facebook partners with Nortek Air Solutions in sustainable data center cooling systems,'' Nortek, 2024. [Online]. Available: \url{https://www.nortekdatacenter.com/facebook-partners-with-nortek-air-solutions-in-sustainable-data-center-cooling-systems/}

\bibitem{ref8}
``Under the sea: Microsoft tests a datacenter that’s quick to deploy and could provide internet connectivity for years,'' Microsoft News, 2025. [Online]. Available: \url{https://news.microsoft.com/features/under-the-sea-microsoft-tests-a-datacenter-thats-quick-to-deploy-could-provide-internet-connectivity-for-years/}

\bibitem{ref9}
J.~Banta, ``StatePoint Liquid Cooling,'' Facebook Engineering, Jun. 5, 2018. [Online]. Available: \url{https://engineering.fb.com/2018/06/05/data-center-engineering/statepoint-liquid-cooling/}

\end{thebibliography}
\end{document}